\documentclass{ragtime} 

\title[Missing bright red giants in the Galactic center   \,\,\,\,\,\,\,\,\,\,\,\, \,\,\,\,       ]{Missing bright red giants in the Galactic center: A fingerprint of its once active state?}

\usepackage{times}

\usepackage[czech,english]{babel}
\usepackage[utf8]{inputenc}
\usepackage{natbib}

\usepackage{float}
\author[M. Zaja\v{c}ek et al.]
       {Michal Zajaček\at[]{1a}, 
        Anabella Araudo\at[]{2,3},                               
        Vladimír Karas\at[]{3},\splitauthors   
        Bo\.{z}ena Czerny\at[]{1},
        Andreas Eckart\at[]{4,5},
        Petra Suková\at[]{3},\splitauthors 
        Marcel \v{S}tolc\at[]{3} and Vojt\v{e}ch Witzany\at[]{6}\\
       \ins{1}Center for Theoretical Physics, Polish Academy of Sciences,\splitins[1]Al. Lotników 32/46, 02-668 Warsaw, Poland\\
       \ins{2} ELI Beamlines, Institute of Physics, Czech Academy of Sciences, \splitins[2] CZ-25241 Doln\'i B\v{r}e\v{z}any, Czech Republic\\
        \ins{3}Astronomical Institute of the Czech Academy of Sciences,\splitins[3]
        Boční II 1401, CZ-14100 Prague, Czech Republic
        \\
        \ins{4} I. Physikalisches Institut der Universit\"at zu K\"oln,\splitins[4] Z\"ulpicher Strasse 77, D-50937 K\"oln, Germany\\
        \ins{5} Max-Planck-Institut f\"ur Radioastronomie (MPIfR),\splitins[5] Auf dem H\"ugel 69, D-53121 Bonn, Germany\\
        \ins{6} School of Mathematics and Statistics, University College Dublin,\splitins[6] Belfield, Dublin 4, D04 V1W8, Ireland\\
        \ins{a}\Email{zajacek@cft.edu.pl}} 

\newcommand{\apj}{ApJ}
\newcommand{\apjl}{ApJL}

\newcommand{\mnras}{MNRAS}

\newcommand{\nat}{Nature}

\newcommand{\aap}{A\&A}

\newcommand{\araa}{ARA\&A}

\newcommand{\physrep}{Physics Reports}
\newcommand{\actaa}{Acta Astronomica}

\begin{document}

\begin{abstract}
 In the Galactic center nuclear star cluster, bright late-type stars exhibit a flat or even a decreasing surface-brightness profile, while fainter late-type stars maintain a cusp-like profile. Historically, the lack of red giants in the Galactic center was discovered via the drop in the strength of the CO absorption bandhead by Kris Sellgren et al. (1990), later followed by the stellar number counts based on the high angular resolution near-infrared observations. Several mechanisms were put forward that could have led to the preferential depletion of bright red giants: star-star collisions, tidal stripping, star-accretion disc collisions, or an infall of a massive cluster or a secondary black hole. Here we propose a novel scenario for the bright red-giant depletion based on the collisions between red giants and the nuclear jet, which was likely active in the Galactic center a few million years ago and could have led to the formation of the large-scale $\gamma$-ray Fermi bubbles. The process of the jet-induced ablation of red giants appears to be most efficient within $\sim 0.04\,{\rm pc}$ (S-cluster), while at larger distances it was complemented by star--accretion disc collisions and at smaller scales, tidal stripping operated. These three mechanisms likely operated simultaneously and created an apparent core of late-type stars within $\sim 0.5\,{\rm pc}$.
\end{abstract}


\section{Introduction}
\label{intro}
The analysis of the intergrated diffuse starlight at $2.3\,{\rm \mu m}$ of the Galactic center region within $\sim 1.2\,{\rm pc}$ revealed a drop in the CO absorption bandhead strength inside $\sim 0.6\,{\rm pc}$ \citep{1990ApJ...359..112S}. Since CO molecules are present in the extended atmospheres of late-type stars, this discovery indicated the missing red-giant problem. Later on, with the development of the adaptive optics technology, individual sources were detected, initially up to magnitude 16 in the $K_{\rm s}$-band, now even up to magnitude 19. This allowed to construct surface-density distributions of late-type and early-type stars. \citet{2009A&A...499..483B} found that late-type stars have a flat surface-density distribution, while young OB stars have a cusp-like distribution in the same region. Recently, \citet{2018A&A...609A..26G} studied the surface density distribution of late-type stars up to the observed magnitudes of $K_{\rm s}=18$ mag. They recovered the previous findings of the core-like distribution for giant late-type stars in the range $12.5-16$ mag, while the faint stars with $K_{\rm s}\approx 18$ mag exhibit a single power-law 3D distribution with the slope of $\gamma\simeq 1.4$ (the surface density slope is $\Gamma\approx \gamma-1=0.4$). \citet{2018A&A...609A..26G} estimate that $\sim 100$ bright giants could be missing within the projected distance of $\sim 0.3\,{\rm pc}$. An independent analysis by \citet{2019ApJ...872L..15H} confirms these findings. They can also recover a single power-law surface-density distribution (``cusp'') of faint giants ($K_{\rm s}<17$ mag) with the projected power-law index of $\Gamma\simeq 0.34$, while they estimate that about 4-5 bright giants appear to be missing within the S-cluster region ($\sim 0.04\,{\rm pc}$). In addition, they constrain the least and the most extended atmospheres of late-type stars in the range between $4$ and $30\,R_{\odot}$ located within $\sim 0.02\,{\rm pc}$. 

By constructing the $K$-band luminosity function (KLF) of late-type stars and by fitting theoretical luminosity functions to it, \citet{2020A&A...641A.102S} inferred that $\sim 80\%$ of the stellar mass of the Nuclear Star Cluster (NSC) formed 10 Gyr or earlier. This episode was followed by a quiescent phase, and another $\sim 15\%$ formed 5 Gyr ago. The remaining few percent could have formed within the last 100 Myr. This implies that the star-formation in the NSC is rather episodic \citep{2011ApJ...741..108P} and most of the stellar mass is old.

In the Galactic center, the two-body relaxation time is of the order of $1\,{\rm Gyr}$,
    \begin{align}
      \tau_{\rm relax}&= \frac{0.34\sigma_{\star}^3}{G^2m_{\star}\rho_{\star}\log{\Lambda}}\sim\,\notag\\
      &\sim  1.8\times 10^9\left(\frac{\sigma_{\star}}{10^2\,{\rm km\,s^{-1}}} \right)^3 \left(\frac{m_{\star}}{1\,M_{\odot}} \right)^{-1} \left(\frac{\rho_{\star}}{10^6\,M_{\odot}\,{\rm pc^{-3}}} \right)^{-1}\,{\rm yr}\,,\label{eq_2body_relaxation}
    \end{align}
where we considered the Coulomb logarithm of the order of 10 and the stellar mass density estimate of $\sim 10^6\,M_{\odot}\,{\rm pc^{-3}}$ is based on the enclosed mass as determined by \citet{2009A&A...502...91S}. The relaxation time could further be shortened by a factor of at least 10 due to the presence of massive perturbers \citep{2007ApJ...656..709P}. Given that $\tau_{\rm relax}$ is comparable or even shorter than the formation time of most of late-type stars, the late-type NSC is expected to be relaxed and its 3D number density should follow a single power-law profile similar to the theoretical Bahcall-Wolf cusp \citep[$n_{\rm BW}\propto r^{-3/2}$ for unequal stellar masses according to \citeauthor{1977ApJ...216..883B}, \citeyear{1977ApJ...216..883B} and $n_{\rm \bullet}\propto r^{-2}$ for stellar black holes; see also][ for reviews]{2005PhR...419...65A, 2017ARA&A..55...17A}. \citet{2020A&A...641A.102S} confirm in their analysis that late-type stars in all magnitude bins follow a single power-law surface-density profile, except for the brightest stars with the observed K-band magnitude in the range of $15-13$ mag, which exhibit a flat to a decreasing surface-density profile that can be described by a broken power law.

To explain this apparent paradox of missing bright red giants, several mechanisms have been proposed based on their preferential effect on bigger stars with more extended, loosely-bound envelopes. Below we list the main proposed mechanisms: 
\begin{itemize}
  \item tidal disruption of red giants and tidal stripping of their envelopes by the supermassive black hole \citep[SMBH; ][]{1975Natur.254..295H,1988Natur.333..523R,2014ApJ...788...99B,2020MNRAS.493L.120K},
  \item red giant--accretion disc (clump) collisions \citep{1996ApJ...470..237A,2014ApJ...781L..18A,2019arXiv191004774A,2016ApJ...823..155K},
  \item collisions of red giants with field stars and compact remnants \citep{1989IAUS..136..543P,1990ApJ...359..112S,1993ApJ...408..496M,1996ApJ...472..153G,2009MNRAS.393.1016D}
  \item mass segregation effects: the infall of a secondary massive black hole \citep{2006MNRAS.372..174B,2006ApJ...648..890M} or the infall of a massive cluster \citep{2003ApJ...597..312K,2009MNRAS.399..141E,2012ApJ...750..111A} or the dynamical segregation of stellar black holes \citep{1993ApJ...408..496M},
  \item central luminosity source as a source of ionizing radiation \citep{1990ApJ...359..112S}. 
\end{itemize}

In the following, we present a novel scenario reminiscent of the last scenario that includes a central luminosity source that photoionizes molecular content in the large envelopes of red giants, including the CO molecule \citep{1990ApJ...359..112S}. However, instead of photoionization, we focus on the possibility of the mechanical ablation of large red-giant atmospheres by a nuclear jet, which could have been significantly more active in the Galactic center a few million years ago. The observed $\gamma$-ray Fermi \citep{2010ApJ...724.1044S} and radio bubbles \citep{2019Natur.573..235H} as well as the X-ray chimneys \citep{2019Natur.567..347P} and optical ionization cones \citep{2019ApJ...886...45B} could be its fingerprints. We describe the model and analyze its consequences in detail in \citet{2020ApJ...903..140Z}. In this contribution, we summarize the main concepts and subsequently, we compare different scenarios of the red giant depletion and how they can complement each other on different spatial scales from Sgr~A*.

\section{Model description}
\label{model}

The basic assumption in our model is the active jet phase of Sgr~A*. Recently, multiwavelength evidence has been accumulated for the presence of bipolar cones -- these include $\gamma$-ray Fermi bubbles \citep{2010ApJ...724.1044S}, radio bubbles \citep{2019Natur.573..235H}, X-ray chimneys \citep{2019Natur.567..347P}, and optical large-scale ionization cones \citep{2019ApJ...886...45B}. Their overall energy content appears to be consistent with the active jet and/or the nuclear disc outflows with the kinetic luminosity of $L_{\rm j}=2.3^{+5.1}_{-0.9} \times 10^{42}\,{\rm erg\,s^{-1}}$ \citep{2016ApJ...829....9M}. On the other hand, a nuclear starburst appears to be inconsistent with the bubble energetics by about a factor of 100 \citep{2003ApJ...582..246B,2019ApJ...886...45B}. \citet{2012ApJ...756..181G} simulated the formation of the Fermi bubbles by an active jet and the basic energetics and properties could be explained by the jet duration of $t_{\rm jet}\sim 0.1-0.5\,{\rm Myr}$. Taking into account the total energy content of ionization cones of $E_{\rm cone}\sim 10^{56}-10^{57}\,{\rm erg}$ \citep{2019ApJ...886...45B}, the jet kinetic luminosity can be estimated as $L_{\rm j}\sim E_{\rm cone}/t_{\rm jet}\sim 6.3 \times 10^{42}- 3.2 \times 10^{44}\,{\rm erg\,s^{-1}}$. This is in agreement with the upper limit given by the Eddington luminosity of Sgr~A*,
\begin{equation}
    L_{\rm Edd}=5\times 10^{44}\left(\frac{M_{\bullet}}{4\times 10^6\,M_{\odot}} \right)\,{\rm erg\,s^{-1}}\,.
    \label{eq_eddington}
\end{equation}
The active jet during the Galactic center Seyfert phase was dominated by the kinetic pressure \citep{2012ApJ...756..181G}. The kinetic luminosity can be estimated using the conversion efficiency $\eta_{\rm j}$ as $L_{\rm j}=\eta_{\rm j}L_{\rm acc}\lesssim \eta_{\rm j} L_{\rm Edd}$, where $L_{\rm acc}$ is the bolometric accretion luminosity. Since $\eta_{\rm j}<0.7$ for most radio galaxies \citep{2008ApJ...685..828I}, this yields the upper limit of $L_{\rm j}<3.5 \times 10^{44}\,{\rm erg\,s^{-1}}$. In the following, we explore the jet-star interaction for $L_{\rm j}=10^{41}-10^{44}\,{\rm erg\,s^{-1}}$, where the lower limit was estimated for the current quiescent phase of Sgr~A* \citep{2012ApJ...758L..11Y}.

\begin{figure}
    \centering
    \includegraphics[width=0.7\textwidth]{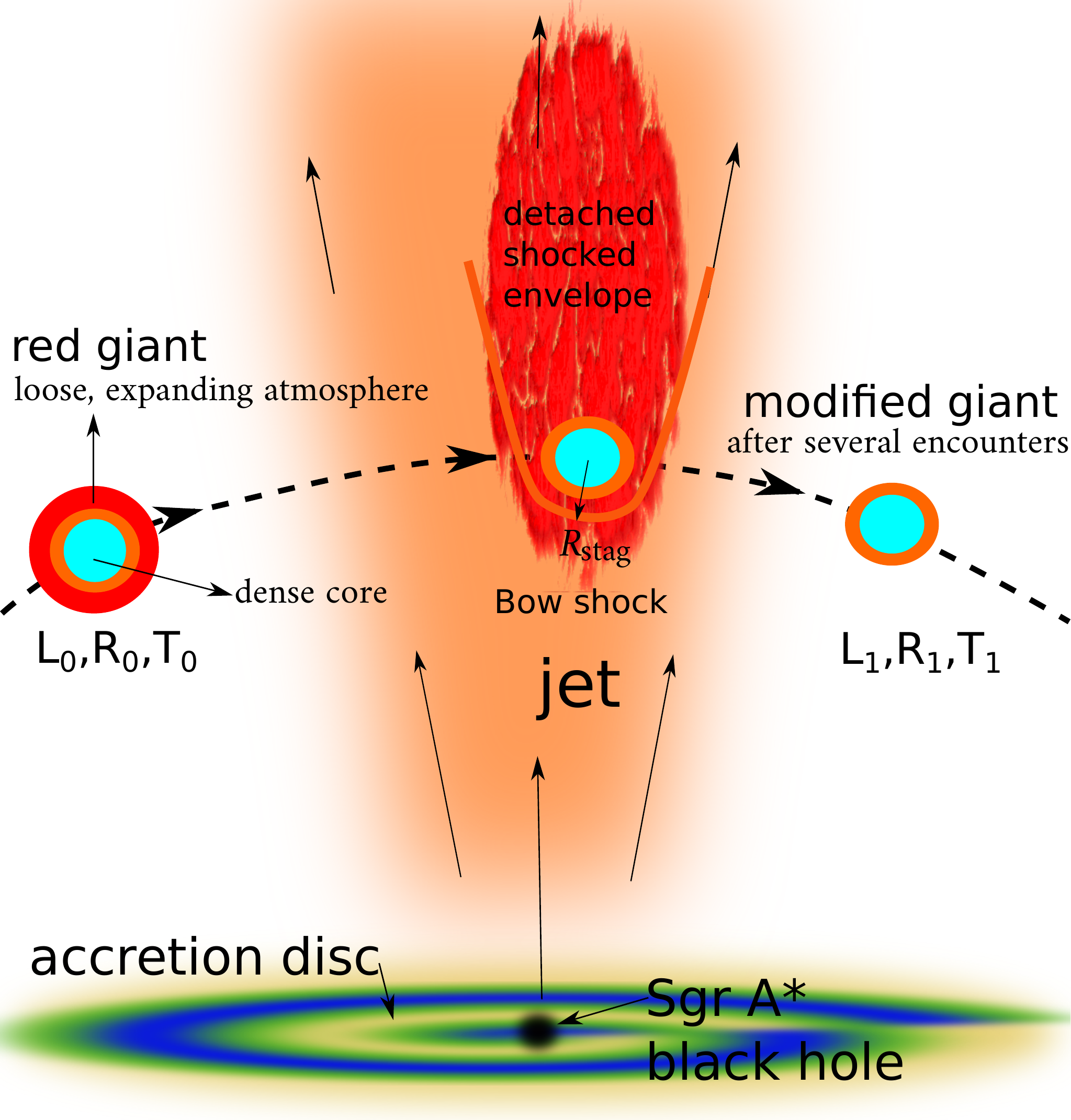}
    \caption{Illustration of the red giant as it crosses the jet during the active phase of Sgr~A* a few million years ago. A red giant consists of a dense core surrounded by a loose and expanding atmosphere. While crossing the jet, its outer layers get ablated, which results in a modified appearance -- higher effective temperature and a drop in near-infrared luminosity after repetitive encounters. For further details, see \citet{2020ApJ...903..140Z}.}
    \label{fig_illustration}
\end{figure}

During the Seyfert phase of Sgr~A* that occurred $3.5 \pm 1$ Myr ago \citep{2019ApJ...886...45B}, most of the NSC late-type stars were certainly present since 80\% of the stellar mass formed at least 10 Gyr ago \citep{2020A&A...641A.102S}. Given this setup, it is straightforward to invoke a scenario where red giants, supergiants, and asymptotic giant-branch stars cross the jet. This scenario was studied extensively in relation to the non-thermal emission of jetted active galactic nuclei \citep[AGN; see e.g.][]{Barkov_10,2012ApJ...749..119B,2013MNRAS.436.3626A,2017A&A...606A..40P}. Here we focus instead on the expected impact of the repetitive jet-star interactions on the visual appearance of red giants, mainly in the near-infrared domain, see Fig.~\ref{fig_illustration} for illustration. Even in the current quiescent state of Sgr~A*, there is evidence for the interaction of wind-blowing stellar objects with the ambient wind or even a low surface-brightness jet \citep{2020MNRAS.499.3909Y}. Clear examples are comet-shaped sources X3, X7 \citep{2010A&A...521A..13M}, and the bow-shock source X8 \citep{2019A&A...624A..97P} located in the so-called mini-cavity. 

\begin{figure}
    \centering
    \includegraphics[width=\textwidth]{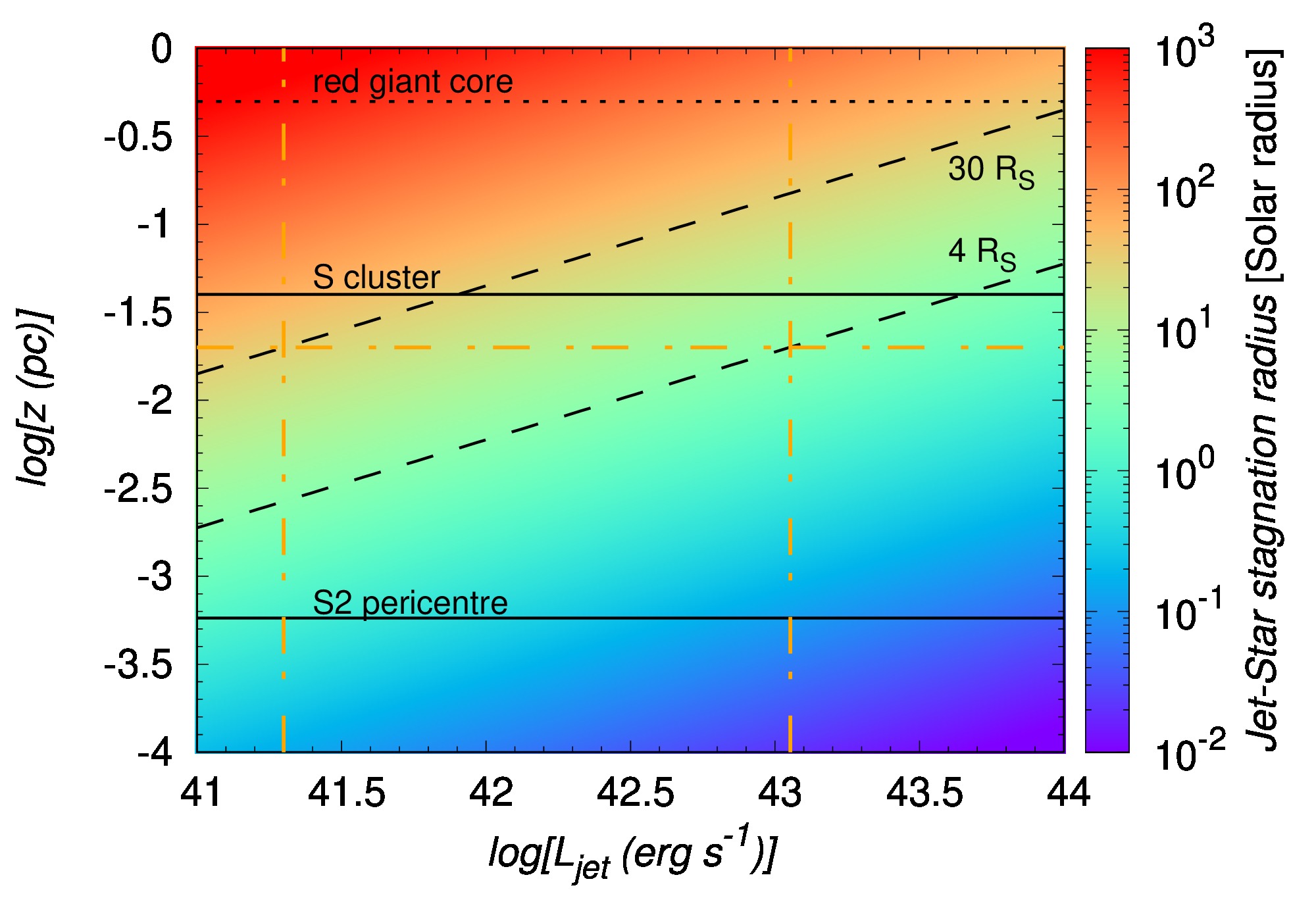}
    \caption{Stagnation radius $R_{\rm stag}/R_{\odot}$ as a function of the jet luminosity (in ${\rm erg\,s^{-1}}$) and of the distance from Sgr~A* (in parsecs). Two horizontal black lines mark the extent of the S cluster -- from the S2 pericenter up to $0.04$ pc (or 1''). Two dashed black lines depict $R_{\rm stag}$ equal to 30 (top) and 4 Solar radii (bottom), respectively. To instruct the reader, we also plot the vertical dot-dashed lines that mark the jet kinetic luminosities, which would result in the atmosphere ablation at 30 $R_{\odot}$ (left) and 4 $R_{\odot}$ (right) at the distance of $z=0.02\,{\rm pc}$. The horizontal dotted black line depicts the approximate length-scale of the red-giant core at $0.5\,{\rm pc}$ \citep[see e.g.][]{1990ApJ...359..112S}.}
    \label{fig_stagnation_radius}
\end{figure}

The basic length-scale that determines where the red-giant envelope is truncated is given by the stagnation radius $R_{\rm stag}$, where the stellar-wind pressure $P_{\rm sw}$ is comparable to the jet kinetic pressure $P_{\rm j}$, which leads to
\begin{align}
  R_{\rm stag}&=z\tan{\theta}\sqrt{\frac{\dot{m}_{\rm w} v_{\rm w} c}{4L_{\rm j}}}= \notag\\
   &=27\left(\frac{z}{0.04\,{\rm pc}}\right)\left(\frac{\dot{m}_{{\rm w}}}{10^{-8}\,M_{\odot}{\rm yr^{-1}}}\right)^\frac{1}{2}\left(\frac{v_{\rm w}}{10\,{\rm km\,s^{-1}}}\right)^\frac{1}{2} \left(\frac{L_{\rm j}}{10^{42}\,{\rm erg\,s^{-1}}}\right)^{-\frac{1}{2}}\,R_{\odot},
   \label{eq_stagnation_radius}
\end{align}
where $z$ is the distance of the star from Sgr~A*, $\dot{m}_{\rm w}$ is the stellar mass-loss rate, $v_{\rm w}$ is the terminal stellar-wind velocity. The half-opening angle $\theta$ is set to $12.5^{\circ}$ \citep[see][ for the opening angle estimate for Sgr~A*]{2013ApJ...779..154L}. The stellar parameters --  $\dot{m}_{\rm w}$ and $v_{\rm w}$ -- are scaled to the typical values for red giants \citep{1987IAUS..122..307R}. According to Eq.~\eqref{eq_stagnation_radius}, the typical stagnation radius in the S cluster region ($\sim 0.04\,{\rm pc}$) is comparable to the largest atmosphere radius of $30\,R_{\odot}$ inferred by \citet{2019ApJ...872L..15H}. In Fig.~\ref{fig_stagnation_radius}, we plot the stagnation radius in Solar radii with respect to the expected jet luminosity and the distance from Sgr~A* (in parsecs). Within the S cluster, the stagnation radius can reach 30$\,R_{\odot}$ up to a few Solar radii only, depending on the exact location and the jet luminosity. These values of $R_{\rm stag}$ generally reach below the atmosphere of larger red giants.

The number of encounters between the jet and the red giant is expected to be at least of the order of 1000 since the orbital timescale $P_{\rm orb}$ is much smaller than $t_{\rm jet}$. Once a red giant enters the jet, it will continue to cross it during subsequent orbits during the jet lifetime. In principle, the vector resonant relaxation and/or the jet precession could cause that the interaction halts. For the following estimates, we assume that these processes take place on longer timescales than the jet lifetime \citep[see also][ for a detailed discussion of these effects]{2020ApJ...903..140Z}. The number of red giant--jet encounters is then approximately,
\begin{equation}
     n_{\rm cross} = 2\frac{t_{\rm jet}}{P_{\rm orb}} \sim 2\times 10^4  \left(\frac{t_{\rm jet}}{0.5\,{\rm Myr}}\right)\left(\frac{M_{\bullet}}{4\times10^6 \,M_{\odot}}\right)^\frac{1}{2}\left(\frac{z}{0.01\,{\rm pc}}\right)^{-\frac{3}{2}}\,.
     \label{eq_number_crossing}
\end{equation}
Within the S cluster, the number of encounters reaches $n_{\rm cross}\simeq 1.4 \times 10^6$ at the S2 pericentre ($r_{\rm p}\sim 0.58\,{\rm mpc}$) and goes down to $n_{\rm cross}\simeq 2500$ at $0.04\,{\rm pc}$. These estimates represent upper limits since during the jet existence the vector resonant relaxation operates that changes the orbital inclination and the star may leave the collisional orbit before the jet ceases its activity. In addition, the same can occur due to the potential jet precession (caused by a secondary black hole or the Lense-Thirring effect) that can change the jet direction on the timescales of tens of years for some sources, such as OJ287 \citep{2018MNRAS.478.3199B}.

During one encounter, the mass removal from the red giant of radius $R_{\star}$ can be estimated from the balance of jet kinetic force and the gravitational force acting on the shell to be removed, $P_{\rm j}\pi R_{\star}^2 \simeq Gm_{\star} \Delta M_1/R_{\star}^2$, from which follows,
\begin{align}
 \frac{\Delta M_1}{M_{\odot}} &\approx 4 \times 10^{-10} \left(\frac{L_{\rm j}}{10^{42}\,{\rm erg\,s^{-1}}}\right)\left(\frac{R_{\star}}{100\,R_{\odot}}\right)^4 \left(\frac{z}{0.04\,{\rm pc}} \right)^{-2}\left(\frac{\theta}{0.22} \right)^{-2} \left(\frac{m_{\star}}{M_{\odot}} \right)^{-1}\,.
 \label{eq_mass_removal_onetime}
\end{align}
The mass removed during $n_{\rm cross}$ encounters can be estimated simply as $\Delta M_{\rm cross}\simeq n_{\rm cross} \Delta M_1$, which yields,
\begin{align}
  \Delta M_{\rm cross} &\sim n_{\rm cross} \Delta M_1\approx \notag\\
           &\approx 10^{-4} \left(\frac{L_{\rm j}}{10^{42}\,{\rm erg\,s^{-1}}}\right)\left(\frac{R_{\star}}{100\,R_{\odot}}\right)^4 \left(\frac{z}{0.01\,{\rm pc}} \right)^{-\frac{7}{2}}\times \notag \\
    & \left(\frac{\theta}{0.22} \right)^{-2} \left(\frac{m_{\star}}{M_{\odot}} \right)^{-1}\left(\frac{t_{\rm jet}}{0.5\,{\rm Myr}}\right)\left(\frac{M_{\bullet}}{4\times10^6 \,M_{\odot}}\right)^\frac{1}{2}\,M_{\odot}\,.\label{eq_Delta_M_cross}
\end{align}
Since 80\% of the stellar mass formed 10 Gyrs or more ago, red giants in the NSC are expected to have gone through several AGN-like phases. Typical AGN phases are short, of the order of $10^5$ years, while the total growth time of the SMBHs is between $10^7$ and $10^9$ years \citep{2015MNRAS.451.2517S}. This implies at least $n_{\rm AGN}\sim 10^2$ AGN-like events during the Galaxy lifetime. The total removed mass during all active phases then is, $\Delta M_{\rm AGN}\sim n_{\rm AGN} n_{\rm cross} \Delta M_1$, which makes the numerical estimate in Eq.~\eqref{eq_Delta_M_cross} larger by at least two orders of magnitude.

The mass removal from red giants estimated by Eqs.~\eqref{eq_mass_removal_onetime}-\eqref{eq_Delta_M_cross} and the associated impulse can also effect the orbit of late-type stars. The effect of one encounter is typically negligible, but the cumulative effect of several thousand crossings through the jet can noticeably change the dynamics of the largest red giants. This is, however, beyond the scope of the current contribution and the effect will be studied in detail in our future studies.

\begin{figure}
    \centering
    \includegraphics[width=\textwidth]{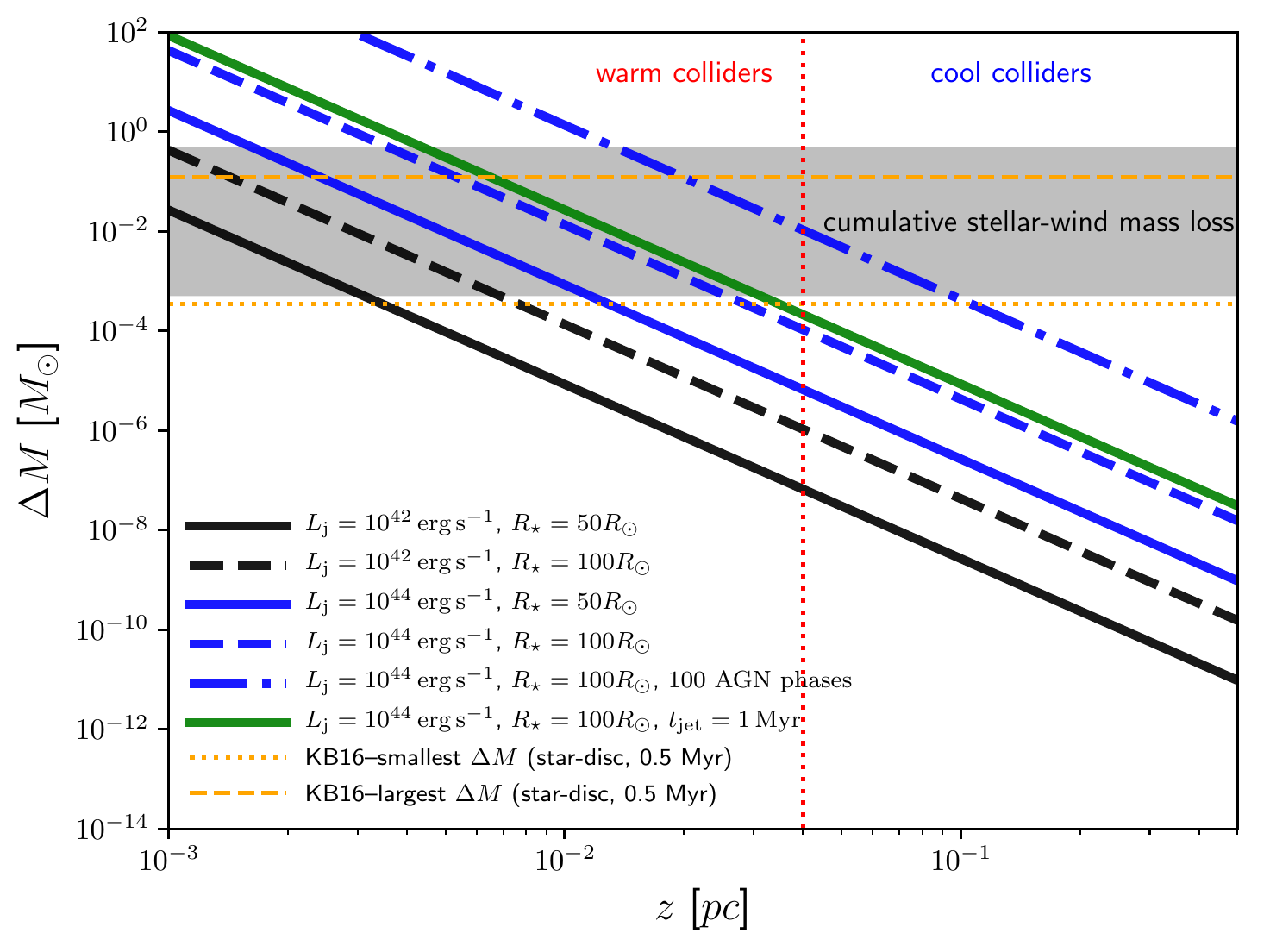}
    \caption{Total mass removed from the red giant atmosphere (in Solar masses) due to the jet ablation as a function of distance from Sgr~A* in parsecs. We plot the cases for $L_{\rm j}=10^{42}\,{\rm erg\,s^{-1}}$ and $L_{\rm j}=10^{44}\,{\rm erg\,s^{-1}}$ and two different atmosphere radii of 50 and 100\,$R_{\odot}$, see the legend (solid and dashed black and blue lines). These cases are nominally calculated for the jet duration of $t_{\rm jet}=0.5\,{\rm Myr}$. In addition, we also include the case for the longer duration of the jet activity, $t_{\rm jet}=1\,{\rm Myr}$ (solid green line). The cumulative mass removal including 100 AGN phases is depicted by a dash-dotted blue line (for the case with $L_{\rm j}=10^{44}\,{\rm erg\,s^{-1}}$ and $R_{\star}=100\,R_{\odot}$). For comparison, we also include the cumulative mass removal for star-disc collisions according to \citet{2016ApJ...823..155K} (dashed and dotted orange lines). Furthermore, the cumulative red-giant (RG) stellar-wind loss during $0.5\,{\rm Myr}$ is shown as a gray rectangle using the observationally inferred mass-loss rates \citep{1987IAUS..122..307R}. The red vertical dotted line depicts the outer radius of the S cluster ($0.04\,{\rm pc}$) and at the same time the approximate dividing radius between the so-called warm and cool colliders \citep{2020ApJ...903..140Z}.}
    \label{fig_total_mass_loss}
\end{figure}

We plot the distance profiles of the removed mass from red giants for different jet luminosities ($L_{\rm j}=10^{42}\,{\rm erg\,s^{-1}}$ and $L_{\rm j}=10^{44}\,{\rm erg\,s^{-1}}$) and atmosphere radii ($R_{\star}=50\,R_{\odot}$ and $R_{\star}=100\,R_{\odot}$) in Fig.~\ref{fig_total_mass_loss}. In addition, we also include the profile for the longer duration of the jet activity ($1\,{\rm Myr}$; solid green line) as well as the cumulative mass removal for red giants going through 100 AGN phases (dash-dotted blue line). For comparison, we also show the cumulative mass removal during star--disc interactions according to \citet{2016ApJ...823..155K} (dashed and dotted horizontal orange lines) assuming that the star-disc mass removal is constant throughout the studied distance range. However, star-disc interactions are expected to remove the mass more efficiently at larger distance scales where the gaseous disc was denser and the Toomre's stability criterion plunged below one because of a larger surface density of the disc. On the other hand, the jet-star interactions are clearly the most efficient in removing the atmosphere mass within the S cluster, where the cumulative mass removal is comparable to the one resulting from standard red-giant stellar winds (shaded gray rectangle) as well as the mass removed during star-disc interactions. This is also enhanced by another effect, which is related to the thermal Kelvin-Helmholtz timescale. In \citet{2020ApJ...903..140Z} we showed that approximately within the S cluster, colliding red giants were not able to cool off before the subsequent collision, which resulted in larger atmosphere cross-sections and hence larger removed mass (warm colliders). Outside the S cluster ($>0.04\,{\rm pc}$), stars were able to cool off because of longer orbital timescales (cool colliders) and their collisional cross-sections were smaller and therefore also the jet-ablation was further reduced by this effect.

Concerning the probability of an encounter between red giants and the jet with a half-opening angle $\theta$, it is clear that for a spherical stellar cluster, not all the stars will interact with the jet at a given moment. In \citet{2020ApJ...903..140Z}, we derive an analytical formula for the mean number of encounters per orbital period in the region with an outer radius $z_{\rm out}$,
\begin{equation}
    \overline{N}_{\rm RG}=\frac{4\pi}{4-\gamma}n_0 z_{0}^{\gamma}\tan{\theta}z_{\rm out}^{3-\gamma}\,,
    \label{eq_number_encounters}
\end{equation}
where $n_0$, $z_0$, and $\gamma$ are the parameters describing 3D number density of late-type stars in the power-law form, $n_{\rm RG}\approx n_0(z/z_0)^{-\gamma}$. For the values inferred by \citet{2018A&A...609A..26G}, $n_0\simeq 52\,{\rm pc^{-3}}$, $z_0\simeq 4.9\,{\rm pc}$, and $\gamma\simeq 1.43$, we obtain $\overline{N}_{\rm RG}\sim 3.5$ within $z_{\rm out}=0.04\,{\rm pc}$ and $\overline{N}_{\rm RG}\sim 82.6$ within $z_{\rm out}=0.3\,{\rm pc}$, which is within uncertainties consistent with the number of missing bright giants at these scales \citep{2018A&A...609A..26G,2019ApJ...872L..15H}. Moreover, the number of encounters given by Eq.~\eqref{eq_number_encounters} is a lower limit as the various dynamical processes (resonant relaxation, jet precession) can effectively increase the interaction volume during the jet lifetime. 

\section{Results: Simulated surface-density profiles}
\label{results}

The Galactic center stellar population can only be studied in detail in the  near-infrared domain, mainly in the $K_{\rm s}$-band at $2.2\,{\rm \mu m}$ \citep[see e.g.][]{2014CQGra..31x4007S}. As the red giant crosses the jet several times, see Fig.~\ref{fig_illustration}, its radius will shrink from $R_0$ to $R_1$, but its bolometric luminosity $L_{\star}=4\pi R_{\star}^2 \sigma T_{\star}^4$ will stay constant as it depends on the core mass \citep{1970AcA....20...47P,1971A&A....13..367R}, which is not affected. This gives us the basic scaling for the effective temperature after the jet-red giant interactions,
\begin{equation}
  T_1=T_0\left(\frac{R_0}{R_1} \right)^{1/2}\,.
  \label{eq_temperature}
\end{equation}
The infrared luminosity $L_{\rm IR}=\pi B_{\rm IR} 4\pi R_{\star}^2$ will decrease with the decrease in the radius. This can be shown by taking the Rayleigh-Jeans approximation, though very crude in the infrared domain, from which $L_{\rm IR}\approx 8\pi^2 (\nu/c)^2 k T_{\star} R_{\star}^2$. The ratio between the post- and the pre-collision infrared luminosity can then be expressed as,
\begin{equation}
    \frac{L^{\rm IR}_1}{L^{\rm IR}_0}\approx \left(\frac{R_1}{R_0} \right)^{3/2}\,.
    \label{eq_ir_luminosity}
\end{equation}
For illustration of the effect, let us consider the red giant with an initial atmosphere radius of $R_0=120\,R_{\odot}$ that is ablated down to $R_1=30\,R_{\odot}$ after several thousand crossings through the jet. From Eq.~\eqref{eq_temperature}, the effective temperature will increase to $T_1=2T_0$, while the infrared luminosity will decrease to $L^{\rm IR}_1=0.125L^{\rm IR}_0$ according to Eq.~\eqref{eq_ir_luminosity} or the magnitude will increase by $2.26$ mag. For an even more profound transition from $R_0=120\,R_{\odot}$ to $R_1=4\,R_{\odot}$, the post-collision values are expected to be $T_1=5.5T_0$, $L_1^{\rm IR}=0.006L_0^{\rm IR}$, and the magnitude increase is $\Delta K=5.54$ mag. These changes can already significantly influence stellar counts in individual magnitude bins. 

\begin{figure}
    \centering
    \includegraphics[width=\textwidth]{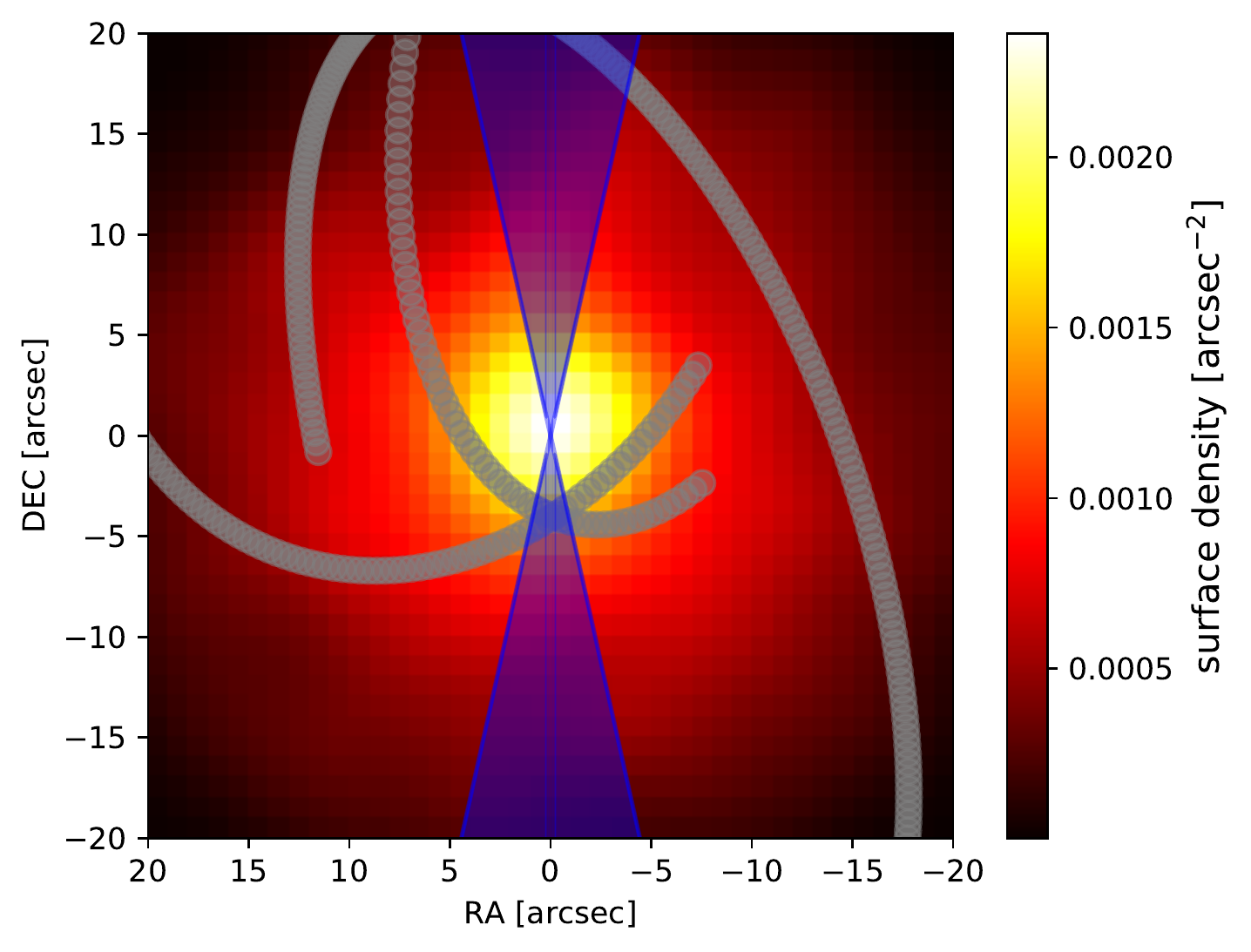}
    \caption{The initial surface-density distribution of the Monte Carlo-generated NSC consisting of 4000 late-type stars, smoothed by the Gaussian kernel on a regular grid of $40\times 40$ points. The blue shaded region represents the jet with a half-opening angle of $12.5^{\circ}$. The three gray streamers represent the three minispiral arms.}
    \label{fig_mock_cluster}
\end{figure}

To better evaluate how the red giant--jet interactions could have affected the surface-density profiles of late-type stars in individual near-infrared magnitude bins, we perform a Monte Carlo simulation by generating a mock NSC. The NSC is assumed to be spherical and described by the volume number density of $n_{\rm RG}\approx n_0(z/z_0)^{-\gamma}$ with $n_0=52\,{\rm pc^{-3}}$, $z_0=4.9\,{\rm pc}$, and $\gamma\sim 1.43$ according to \citet{2018A&A...609A..26G}. This number density profile implies the presence of $\sim 4000$ late-type stars inside one parsec, whose properties were generated using the Monte Carlo approach. The surface density distribution of such a cluster is depicted in Fig.~\ref{fig_mock_cluster} including the jet with a half-opening angle of $\theta=12.5^{\circ}$ and the three minispiral streamers as currently observed are also plotted for a better orientation.

Each star was assigned a mass in the range between $0.08\,M_{\odot}$ and $100\,M_{\odot}$ according to the Kroupa initial mass function \citep[IMF; ][]{2001MNRAS.322..231K}. The Chabrier/Kroupa IMF seems to be consistent with the IMF of the Galactic center late-type stellar population \citep{2011ApJ...741..108P}. Subsequently, we also assigned the core mass to each star. For the purposes of our analysis, we fixed the core-mass fraction to $\mu_{\rm c}/m_{\star}=0.4$, which lies between the value derived from the Sch\"onberg-Chandrasekhar limit and the values expected from the last phase of the stellar evolution, when the white-dwarf core constitutes most of the mass of solar-type stars.  

To calculate the magnitude distribution in the near-infrared $K_{\rm s}$-band, we followed this procedure,
\begin{enumerate}
   \item We calculate the bolometric stellar luminosity and the stellar radius as a function of the core mass, $L_{\star}(\mu_{\rm c})$ and $R_{\star}(\mu_{\rm c})$, according to Eq. (19) in \citet{2020ApJ...903..140Z}.
   \item When the jet is active, we compare $R_{\star}$ of the entering star with $R_{\rm stag}$ calculated using Eq.~\eqref{eq_stagnation_radius}. When $R_{\star}\geq R_{\rm stag}$, we set $R_{\star}=R_{\rm stag}$. For this analysis, we assumed that almost all late-type stars within $0.5\,{\rm pc}$ could have interacted with the jet for at least several hundred times. This would be possible when the vector resonant-relaxation timescale is shorter than the jet lifetime or the jet would be precessing.
   \item The effective temperature is calculated as $T_{\star}=T_{\odot} (L_{\star}/L_{\odot})^{1/4} (R_{\star}/R_{\odot})^{-1/2}$.
   \item From the Planck function $B_{\nu}(T_{\star})$, we calculate the flux density at $K_{\rm s}$-band, $F_{\nu}=\pi B_{\nu}(T_{\star})(R_{\star}/d)^2$, where $d$ is the distance to the Galactic center. The intrinsic (dereddened) apparent magnitude is then calculated as $m_{\rm K}=-2.5\log{(F_{\nu}/653\,{\rm Jy})}$.
\end{enumerate}

\begin{figure}
    \centering
    \includegraphics[width=\textwidth]{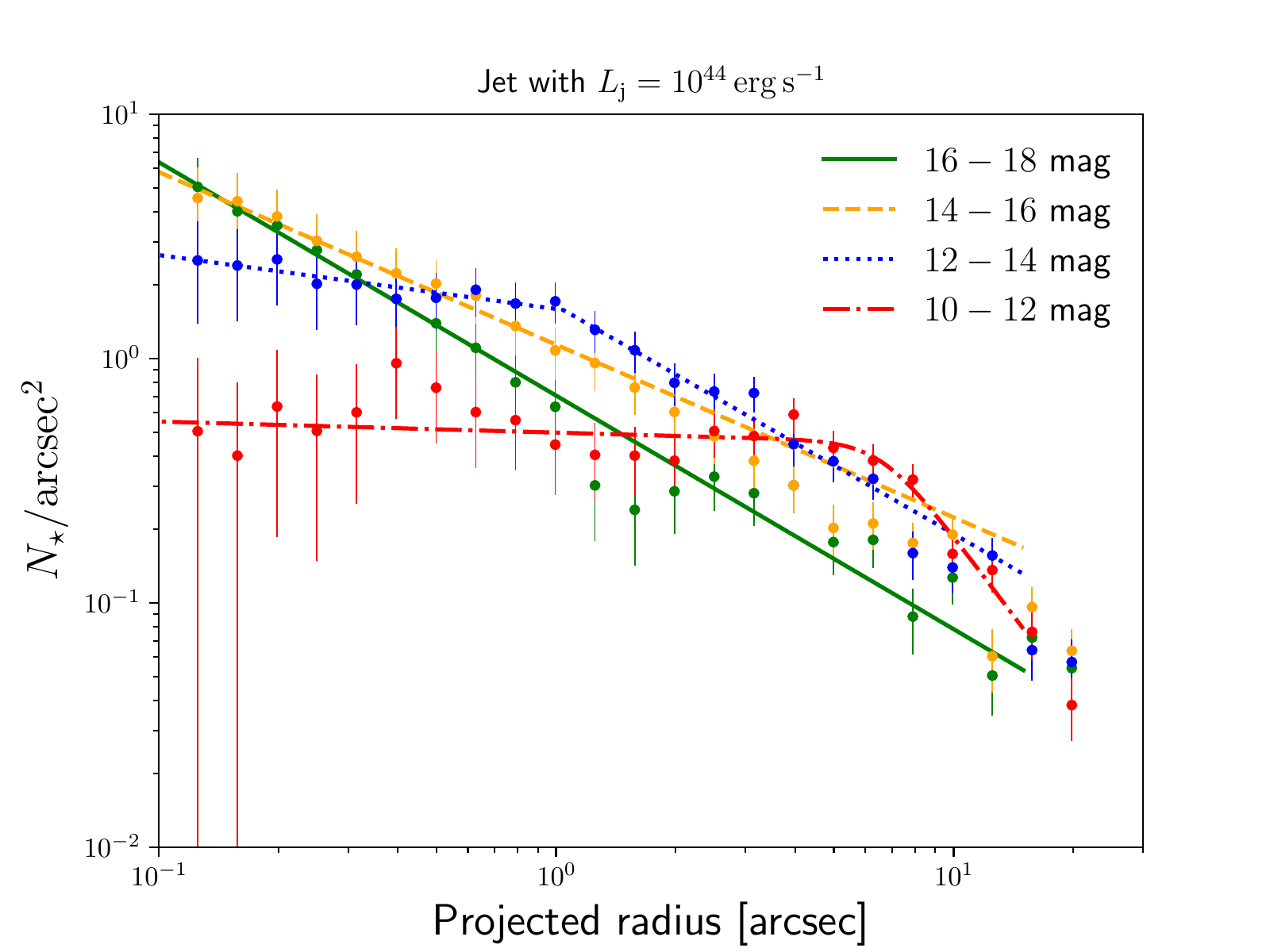}
    \caption{Post-collisional surface-density distributions of late-type stars constructed from an initially cuspy NSC. The jet kinetic luminosity was set to $10^{44}\,{\rm erg\,s^{-1}}$ and all the stars were assumed to interact with the jet, which for the brightest giants led to the ablation of their envelopes. The brighter late-type stars in the magnitude bins 10-12 mag and 12-14 mag exhibit a broken power-law distribution, while the fainter giants (14-16 mag and 16-18 mag) maintain the cusp-like distribution. The magnitude values are treated as intrinsic or dereddened. For the comparison with the observational results, it is necessary to add $\sim 2.5$ mag \citep{2010A&A...511A..18S} for the line-of-sight mean extinction in $K_{\rm s}$-band.}
    \label{fig_surface_densoty_L44}
\end{figure}

In the next step, to estimate the surface-density distributions, we count the number of stars $N_{\star}$ in the concentric annuli with the mean radius $R$ and the width $\Delta R$, from which we estimate the surface density as $\sigma_{\star}=N_{\star}/(2\pi R \Delta R)$ with the uncertainty of $\sqrt{N_{\star}}/(2\pi R \Delta R)$. Then we bin the stars into two-magnitude bins, starting at 18 mag and going down to the brightest stars with 10 mag. Initially, without any jet influence, the surface density distribution across all magnitude bins could be described as a single power-law described as $N(R)=N_0(R/R_0)^{-\Gamma}$, hence as a proper cusp, see Table~\ref{tab_power_law_distributions}. When the jet is switched on with the kinetic luminosity of $L_{\rm j}=10^{44}\,{\rm erg\,s^{-1}}$, the surface-density profile of the brightest giants (10-12 mag, dereddened) becomes flat up to $\sim 0.3\,{\rm pc}$, see Fig.~\ref{fig_surface_densoty_L44}. The flattening inside $0.04\,{\rm pc}$ is also apparent for late-type stars in the 12-14 mag bin (dereddened), while fainter stars with larger magnitudes maintain a cusp-like distribution as the most recent observational studies indicate \citep{2018A&A...609A..26G,2019ApJ...872L..15H,2020A&A...641A.102S}. The surface-density distribution of bright late-type stars can be fitted by a broken power-law function in the form $N(R)=N_0(R/R_{\rm br})^{-\Gamma}[1+(R/R_{\rm br})^{\Delta}]^{(\Gamma-\Gamma_0)/\Delta}$, where $R_{\rm br}$ is the break radius, $\Gamma$ is the slope of the inner cluster part inside $R_{\rm br}$, $\Gamma_0$ is the slope of the outer cluster part, and $\Delta$ describes the sharpness of transition. For both the case without any jet and the case with the jet kinetic luminosity of $10^{44}\,{\rm erg\,s^{-1}}$, we summarize the slopes and the break radii of the fitted power-law functions in Table~\ref{tab_power_law_distributions} for individual magnitude bins.  

\begin{table}[tbh]
    \centering
     \caption{Best-fit slopes and break radii of the power-law distributions that are used to describe the surface-density distributions of the mock NSC at its initial ``cuspy'' stage (no jet), see Fig.~\ref{fig_mock_cluster}, and in the stage after the active phase with the jet, see distributions in Figure~\ref{fig_surface_densoty_L44}. Values are listed for different magnitude bins and two cases of the jet activity: no jet and the jet with the kinetic luminosity of $L_{\rm j}=10^{44}\,{\rm erg\,s^{-1}}$.}
    \begin{tabular}{c|c|c}
    \hline
    \hline 
     Magnitude bin & No jet activity & Jet $L_{\rm j}=10^{44}\,{\rm erg\,s^{-1}}$  \\
     \hline
    18-16 mag  &  single: $\Gamma=0.9$  & single: $\Gamma=1.0$\\
    16-14 mag  &  single: $\Gamma=0.6$  & single: $\Gamma=0.7$\\
    14-12 mag  &  single: $\Gamma=0.6$  & broken: $\Gamma=0.2$, $\Gamma_0=0.9$, $R_{\rm br}=1''$\\
    12-10 mag  & single: $\Gamma=0.7$  & broken: $\Gamma=0.04$, $\Gamma_0=2.2$, $R_{\rm br}=6.6''$\\
    \hline
    \end{tabular}
    \label{tab_power_law_distributions}
\end{table}

\begin{figure}
	\centering
	\includegraphics[width=\textwidth]{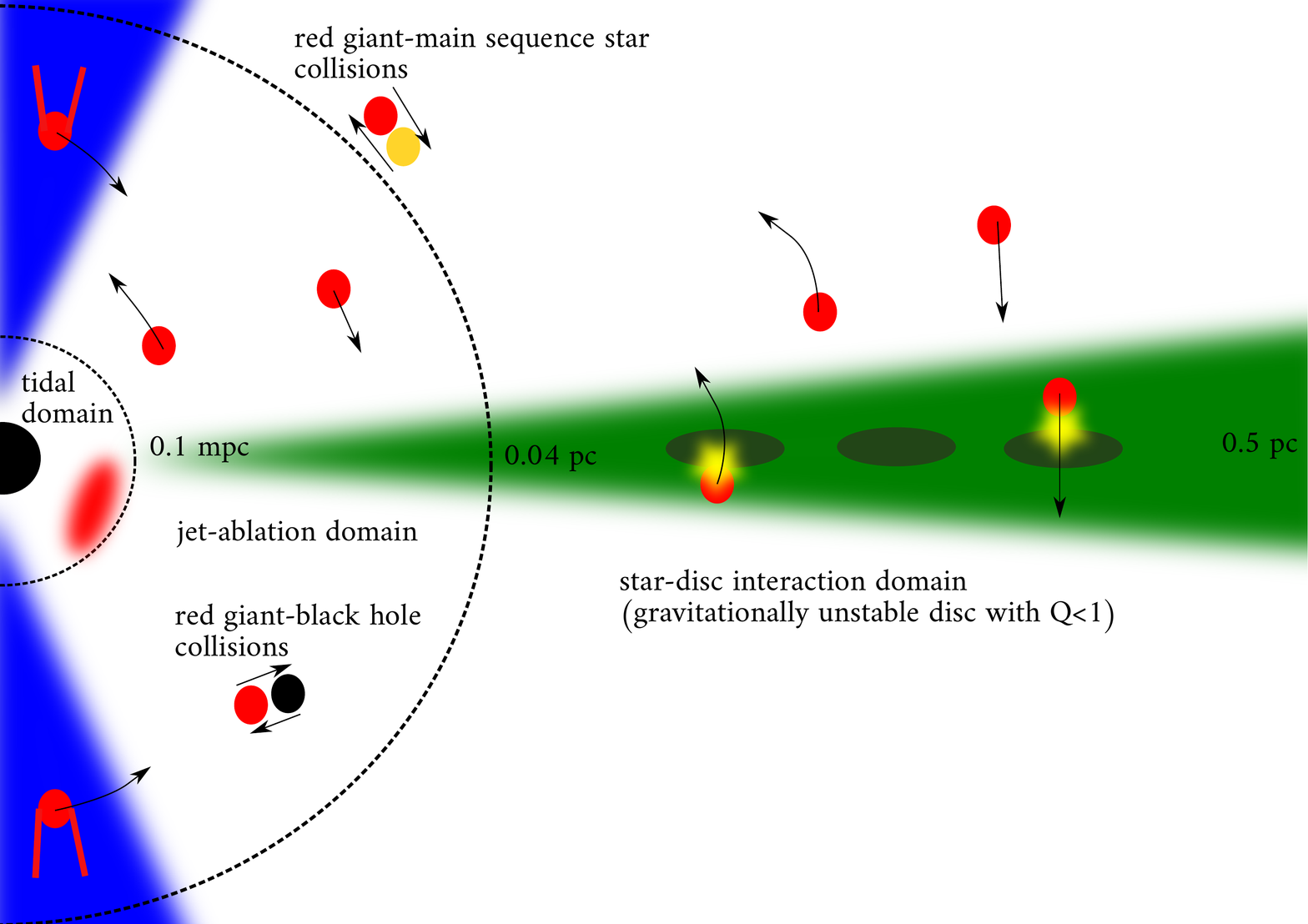}
	\caption{Illustration of the mechanisms that contributed to the depletion of bright red giants in the inner $0.5\,{\rm pc}$ of the Galactic center. Closest to Sgr~A* ($0.1$ mpc), the tidal stripping of the red giant envelopes operated. Inside the inner $0.04\,{\rm pc}$ (S cluster), jet-red giant interactions dominated, while outside $0.04\,{\rm pc}$, the interaction of red giants with the fragmenting gaseous disc was likely the dominant mechanism. Occasional red giant-black hole and red giant-main sequence star collisions further contributed to the depletion throughout the NSC. Distances are not drawn to the scale.}
	\label{fig_unification}
\end{figure}

\section{Discussion and Conclusions}
\label{conclusions}

We presented a novel scenario based on the red giant--nuclear jet interactions to explain the peculiarities of the Milky Way Nuclear Star Cluster -- on one hand, the cusp-like distribution of faint late-type stars, which is expected from the two-body relaxation, and on the other hand, the core-like distribution of bright red giants, which requires a mechanism that preferentially acts on stars with more extended atmospheres. The interaction of red giants with the nuclear jet during active phases of Sgr~A* can naturally explain the observed surface-density features since the jet removes the envelopes of more extended red giants more efficiently from basic principles. The repetitive mechanical removal of the atmosphere material also ensures that the resulting effect on the late-type star is permanent -- its effective temperature will increase, its mass will decrease, and the infrared luminosity will drop. This is an advantage of this model in comparison with the central luminosity source proposed by \citet{1990ApJ...359..112S}, since after the source luminosity drops, the CO molecule could form again in the extended atmosphere because of the lowered ionizing potential. 

In addition, we showed that the jet--red giant interactions are the most efficient in removing the mass inside $0.04\,{\rm pc}$ or the S cluster region, where the removed mass by the jet ablation is comparable to the cumulative mass loss from stellar winds. Hence, the overall mass loss can effectively be doubled in comparison with the standard stellar evolution during the jet existence. The mass removal in the S cluster could also be enhanced by the fact that red giants are expected to be warm colliders in this region, i.e. stars that are not able to cool off before the subsequent collision with the jet. Therefore their envelopes are more puffed up, which increases the collisional cross-sections and the overall mass loss. For bright giants with atmosphere radii of the order of $100\,R_{\odot}$, the increase in the near-infrared magnitude is expected to be $2-6$ mag, which leads to the flattening of their surface-density profile, while they can turn into fainter gaints in the near-infrared after the ablation and make their profiles even more cuspy.  

Outside the S cluster, the efficiency of jet--red giant interactions in terms of the mass removal drops in comparison with star-disc collisions. Therefore, the star-disc collisions likely complemented jet--red giant interactions at larger scales, where the massive gaseous disk fragmented into star-forming clumps with significantly increased density \citep{2003ApJ...590L..33L,2004ApJ...604L..45M}. On the other hand, at much smaller scales of a fraction of a milliparsec, tidal stripping of envelopes operated. Throughout the NSC, occasional collisions of large red giants with stellar black holes and main-sequence stars could partially have contributed to the depletion, but they cannot alone explain the missing brightest giants \citep{2009MNRAS.393.1016D}. We illustrate all of these mechanisms and the regions of their largest efficiency in Fig.~\ref{fig_unification}. They likely all contributed to the observed dearth of bright red giants in the inner $0.5\,{\rm pc}$. 

A limitation of our model is that only a fraction of late-type stars that at a given time cross the jet will suffer the mass removal because of the narrow opening angle of the jet. A wide-angle outflow, e.g. resulting from the accretion disc winds, with a half-opening angle $w\theta$, with $w>1$, could hit more stars. However, this is at the cost of enlarging the stagnation radius by $\sim w$ and diminishing the mass removal by $w^2$.  Hence, a highly-collimated jet is required for the mechanism to work efficiently enough to create an apparent core in the surface-density distribution. Moreover, discussed dynamical processes, such as the coherent resonant relaxation and the jet precession, can considerably enlarge the interaction volume during the active jet phase. Since there are at least $\sim 100$ such phases during the Galaxy lifetime, the total number of affected giants is consistent with the number of missing large giants.

\ack

We would like to thank the organizers of RAGtime 22  (dedicated to Prof. Zden\v{e}k Stuchlík) for the smooth and efficient performance of the conference in the online space during the COVID19 pandemic. MZ thanks Robert ``Ski" Antonucci (UC Santa Barbara) for input and useful comments. 
MZ and BC acknowledge the financial support by the National Science Center, Poland, grant No. 2017/26/A/ST9/00756 (Maestro 9). VK thanks M\v{S}MT grant LTI 17018 for support. We are also grateful to the Czech-Polish mobility program (M\v{S}MT 8J20PL037) for support. In addition, MZ also acknowledges the NAWA financial support under the agreement PPN/WYM/2019/1/00064 to perform a three-month exchange stay at the Astronomical Institute of the Czech Academy of Sciences in Prague. VW was supported by European Union’s Horizon 2020 research and innovation programme under grant agreement No 894881.


\end{document}